\documentclass[10pt, conference]{IEEEtran}
\ifCLASSINFOpdf
  \usepackage[pdftex]{graphicx}
\else
\fi
\usepackage{amsmath}
\usepackage{float}
\usepackage{graphicx}

\begin{document}
%
\title{Programming Bots by Synthesizing \\Natural Language Expressions into API Invocations}




\author{\IEEEauthorblockN{Shayan Zamanirad, Boualem Benatallah, Moshe Chai Barukh, Fabio Casati\IEEEauthorrefmark{1}, and Carlos Rodriguez}
\IEEEauthorblockA{School of Computer Science and Engineering\\
University of New South Wales (UNSW),
Sydney, NSW 2052\\
\{shayanz, boualem, mosheb, crodriguez\}@cse.unsw.edu.au}
\IEEEauthorblockA{\IEEEauthorrefmark{1}University of Trento, Italy / Tomsk Polytechnic University, Russia\\
fabio.casati@unitn.it}}


\maketitle

\begin{abstract}
At present, bots are still in their preliminary stages of development. Many are relatively simple, or developed ad-hoc for a very specific use-case. For this reason, they are typically programmed manually, or utilize machine-learning classifiers to interpret a fixed set of user utterances. In reality, real world conversations with humans require support for dynamically capturing users expressions. Moreover, bots will derive immeasurable value by programming them to invoke APIs for their results. Today, within the Web and Mobile development community, complex applications are being stringed together with a few lines of code -- all made possible by APIs. Yet, developers today are not as empowered to program bots in much the same way. To overcome this, we introduce \textit{BotBase}, a bot programming platform that dynamically synthesizes natural language user expressions into API invocations. Our solution is two faceted: Firstly, we construct an API knowledge graph to encode and evolve APIs; secondly, leveraging the above we apply techniques in NLP, ML and Entity Recognition to perform the required synthesis from natural language user expressions into API calls.
 
\end{abstract}


%
\IEEEpeerreviewmaketitle

\section{Introduction}

The API economy is in full swing. Applications are rapidly developed by composing APIs. Companies increasingly structure their development and internal systems in terms of APIs, and business strategies for many companies revolve around APIs. This novel economy facilitates the creation of new business models and opens many revenue opportunities \cite{api-economy}. Technological advances, the cloud, awareness at the CxO level (starting from the famous memo by Jef Bezos \cite{hernandez2012jeff}), and the increased need for business agility are making applications based on composing APIs mainstream. The ``Internet of Things'' is pushing this trend even more by providing API access to all sort of devices, so that we can use APIs not only to place an order but also to make coffee.

This paper leverages the API economy in combination with advances in bots technology to facilitate the development of intuitive computing solutions that connect user needs, expressed in natural language, with invocations of the APIs that can fulfill these needs. Intuitive computing envision a collaboration between users and ``an intelligent digital assistant that models user intent, and that suggests or carries out actions likely to satisfy that intent'' \cite{int-comp}. Developing applications that enable such kind of collaboration today is facilitated by the many bot builders frameworks and systems provided by nearly every big IT player such as Facebook \cite{witai}, Google \cite{apiai}, Microsoft \cite{botframework}, Amazon \cite{lex}, IBM Watson \cite{ibmconv} and more.

Messaging bots could in fact, be well set as the viable alternative or even successor of mobile apps \cite{bot-rise,bot-over}. Text messaging itself are already embedded in many apps and used by over 3 billion people, with this number is increasing rapidly \cite{bot-rise}. Messaging bots look and feel not much different to chatting with a friend via instant messaging. There is no need to learn, understand or navigate disparate interfaces or languages -- yet they carry the potential of being programmed to automate conversations, transactions or workflows \cite{bot-over}. 

However, at present, bot building systems while useful in detecting the user's intent, still require significant
development and configuration work for each usage scenario, and they hardcode the dependency with the APIs to be invoked. In other words, today, for each class of functionality we want to enable through a bot, and for each API, we need to hardcode the logic that interacts with the user to identify their intent and obtain the desired values for the specific API parameters. Bot building framework have no knowledge of these APIs: they support the identification of specific intents and parameters and then they can be configured to make a REST call. APIs are therefore external to the bot builder and is completely outside of the process, only appearing at the end in an hardcoded fashion.

Our vision consists of making APIs first class citizens of bot builders. We aim at synthesizing natural language expression, and at dynamically determining which API to invoke based on our understanding of the users' intent and on the knowledge over an API knowledge graph that describes what the methods do and how they can be invoked. The vision we set forth in this paper is that of users being able to talk with assistants (as some of us do every day with Siri, Alexa or Cortana) and, with the help of a knowledge of APIs modeled via a knowledge graph and built incrementally, dynamically identify intents, APIs fulfilling the identified intent, and collect from the user the value of the required parameters for invocation. If successful, this can enable a new approach to the development of cognitive services where the ``program'' is built on the fly based on users' requests and available services exposed through APIs.

More specifically, we devise a technique for synthesizing natural language user expressions into concrete API calls by leveraging an API Knowledge Graph (KG) to achieve this. The API KG contains information about APIs, their declarations, expressions, parameters and possible values.

\section{Preliminaries}\label{prelimin}

\vspace{0.15cm}

The following are accepted terminology as related to bot-technology \cite{witai}\cite{apiai}\cite{ibmconv}:

\textbf{\emph{Expression.}} This refers to a natural language user utterance whilst conversing with a bot (e.g. \textit{'what is the weather forecast for tomorrow?'}). Based on the users' expression, the bot recognizes a category such as \textit{'weather\_condition', 'find\_restaurant'}. This refers to the users' intention.

\textbf{\emph{Intent.}} The notion of intent is widely used in conversation systems, and refers to the users' purpose in an expression. For the example above, the underlying intent may be summarized as \textit{'weather\_condition'}. In conversation systems, a fundamental step to answer a users' query is to recognize intention.

\textbf{\emph{Entity.}} A common notion amongst NLP systems, an entity represents a concept specific to domain or context. In the previous example, an expression such as \textit{'what is the weather like tomorrow?'} identified a keyword such as \textit{'tomorrow'}. When reshaped as an entity, we link this to a type such as \textit{time/date}. Formulating entities often requires analyzing the keyword in conjunction with the overall intent.

\textbf{\emph{Conversation.}} In order to execute an action by the bot (e.g. calling an API), all required parameters would need to filled in prior to execution. However, in the event that something is missing, the bot will typically prompt users with more questions to obtain any missing information. This back and forth question-answering between user and bot (and vice versa) is referred to as conversation.

\textbf{\emph{Action.}} This ties a high-level user expression into a concrete execution plan, in order to obtain the required results to satisfy the user's query. For example, we may have an action such as \textit{FindRestaurant}. In many ways, actions are akin to functions in traditional programming languages. Whereby, just like functions, actions also often requires input \textit{parameters}. For example, more specifically the action above could likely be described as \textit{FindRestaurant(type,place,location)}, such as: FindRestaurant(`Italian',`cafe',`Sydney').

\noindent \textbf{Our Approach.} \emph{BotBase} builds upon all the above concepts, plus in our work we propose the notion of \emph{\textbf{Declaration}}, which refers to the expressivity of an API. Moreover, the notion of API and its constituent \emph{declarations}, in our context are synonymous with the traditional notion of \emph{intent}. 

\begin{figure}[h!]
\centering
\includegraphics[scale=0.689]{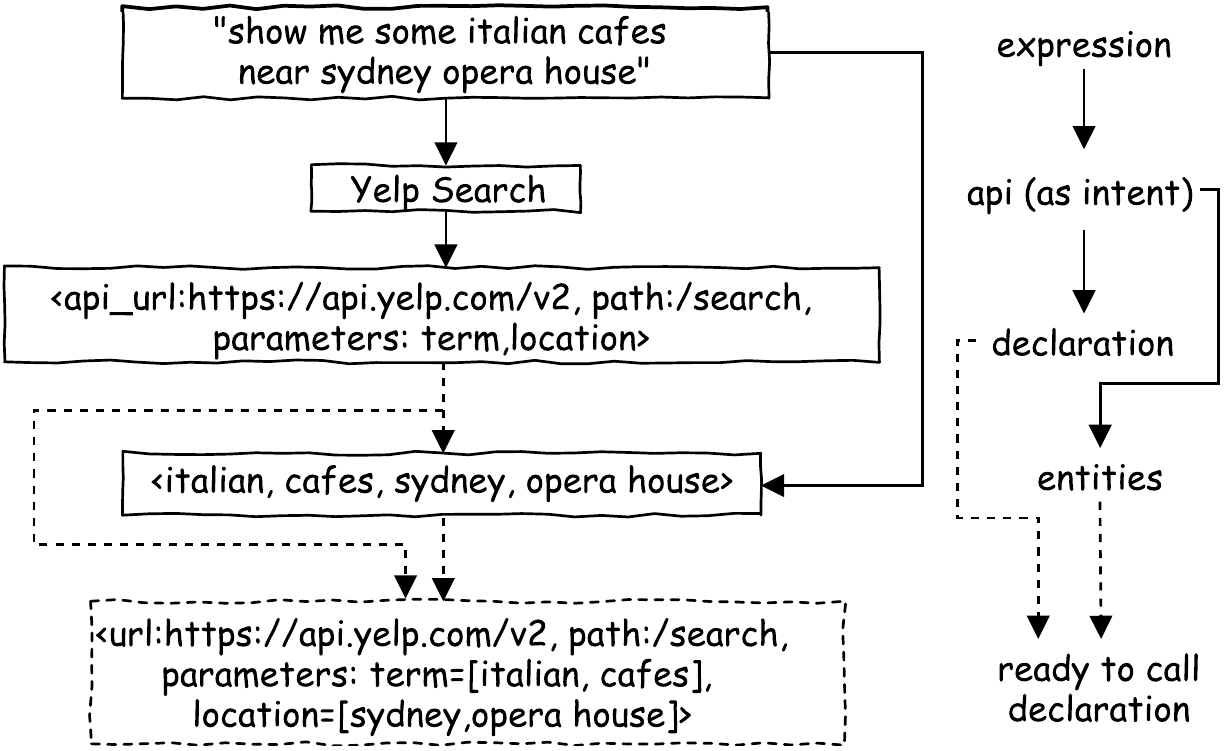}
\vspace{-0.2cm}
\caption{\emph{BotBase} Approach -- An API-oriented Methodology}
\label{fig:decl-exp-mention}
\end{figure}

Figure \ref{fig:decl-exp-mention} illustrates the relationship between expression, APIs, Declarations, Entities and API Calls.

\section{API Knowledge Graph}

\subsection{Knowledge Model}\label{subsect:km}
We propose a lightweight model to capture useful information about APIs (and their related information). Such  knowledge, which may be created incrementally and collectively, will significantly assist bot programming both during development and execution. During development, it is the onus of the developer to properly specify set of expressions (or patterns thereof), possible keywords and the required actions for each. Clearly, specifying this each time is not only tedious but likely to lack full inclusion (e.g. there may be some valid expressions that the developer did not think about). Thus, the value of a knowledge graph in this context would alleviate these challenges. More so, in our proposed work the synthesis technique performs in real-time (with the help of the knowledge-graph), thus also acting during execution time to assist in the bot's processing of the result.

The following are the main entities of our model: \emph{\textbf{API}} is the root and contains a \emph{name}, \emph{description}, a set of descriptive \emph{tags} and a \emph{base\_URI} (e.g. \texttt{api.yelp.com}). An API may have one or more \emph{\textbf{Declarations}} that are supported for this particular API. It contains a type (e.g. \texttt{GET} or \texttt{POST}) and a \emph{path} (e.g. \texttt{/search?term=[term]\&location=[location]}).
Declarations are linked to a set of textual \emph{Sample Expressions}. Declarations may also have one or more \emph{\textbf{Parameters}} (e.g. \textit{term, location}); which may be linked to possible values \emph{\textbf{Para\_Values}} (e.g. \textit{food, San Francisco}). Parameters also has a \emph{required} field to flag if the parameter is optional or not. 

Finally, we define an entity called \textit{Value\_WEValue} that links concrete values with semantically similar words that are derived from our deep learning model. (Note, WE is an acronym for \emph{Word Embedding}.) 

\subsection{Knowledge Acquisition}\label{sec:knowledge-aq}

We employ a combination of two techniques for acquisition and enrichment of new knowledge: Firstly, we harness \emph{crowdsourcing}, and this paradigm enables an incremental and collective approach. Crowdsourced workers such as (i) developers enrich the KG by entering sample expressions at bot training time, (ii) end-users enrich the KG whilst talking with bots. Secondly, we apply \emph{deep learning}, and this paradigm introduces a layer of intelligence and automation. 

\vspace{0.2cm}\noindent\textbf{Word Embedding -- Deep Learning Model.} We use a deep-learning model called Word Embedding\cite{mikolov2013distributed} to generate semantically similar keywords, that relate to existing parameter values. To more the dense, the easier it makes it to more successfully link arbitrary user expressions with API declaration. This is because we would be able to better recognize a match to the API's declaration and parameters, and find an API that would useful to process the expression's intent. 

More specifically, the model we use is Word2vec \cite{mikolov2013distributed}. This model takes as input a corpus and produces a vector-space where each word in the corpus is represented by a vector. Words that are semantically similar are close to each other in the vector space and such distance is typically measured with the help of the cosine similarity metric \cite{Jurafsky:2009:SLP:1214993}. While our model can be trained on any text corpus, in this paper we use a dataset of Wikipedia\footnote{https://dumps.wikimedia.org/} because it is general and it covers a variety of topics. This corpus contains more than 2 billion words. The neural network uses 150 hidden layers; sliding window size of five; skip gram with negative sampling\cite{mikolov2013distributed}; sets the minimum occurrence of a words to fifty (i.e. the word must appear at least fifty times to be inside the training set); and ignores stopwords by using the NLTK\cite{birdnatural} library for preprocessing the corpus. Our trained model contains both unigrams (e.g. car, restaurant, computer) and biagrams\footnote{https://en.wikipedia.org/wiki/Bigram}(e.g. Opera House, New York, ice cream). 

We use our trained model (i) in API KG enrichment process: for a given parameter value (e.g. \textit{"Paris"}), we can add its neighbours (from the word embedding model) such as \textit{"Sydney"}, \textit{"New York"}, \textit{"London"} as another possible values for that parameter (e.g. \textit{"location"}), (ii) in the synthesis process: for an extracted value (e.g. \textit{"Melbourne"}) from user expression, we can link it to the most relevant parameter (e.g. \textit{"location"}) by computing similarity ratio between the extracted value and values already linked to that parameter inside the API KG.

\section{Synthesis of Expression to API}\label{synthexpapi}

\textit{BotBase} is actualized as a set of components, that when placed in orchestration serves to fulfill the fundamental goal of mapping a natural language \emph{user expression} into a concrete {API call}. We refer to this overall process as ``synthesis''. Underlying this process, the components critically relies on the intelligence of the knowledge graph and word embedding model that we proposed in the previous section. This process (or even individual components) could potentially be used at every stage of a bots lifecycle. For example, during \emph{bot development}, a bot developer usually have to select the API and declaration from a pool of APIs. Especially when there are a large number of APIs, this could be much simplified using the KG to help guide the developer select the matching API/s. The developer could input a set of sample user expressions -- and then feed them into the process to discover the API that the synthesis process deems most relevant. Similarly, during \emph{bot training}, developers may enter sample expressions, to allow the bot to learn possible values for the various parameters of the selected API declaration. Finally, during \emph{execution}, of course it serves the fundamental purpose of processing user expressions into API calls.

\begin{figure}
\centering
\includegraphics[scale=0.48]{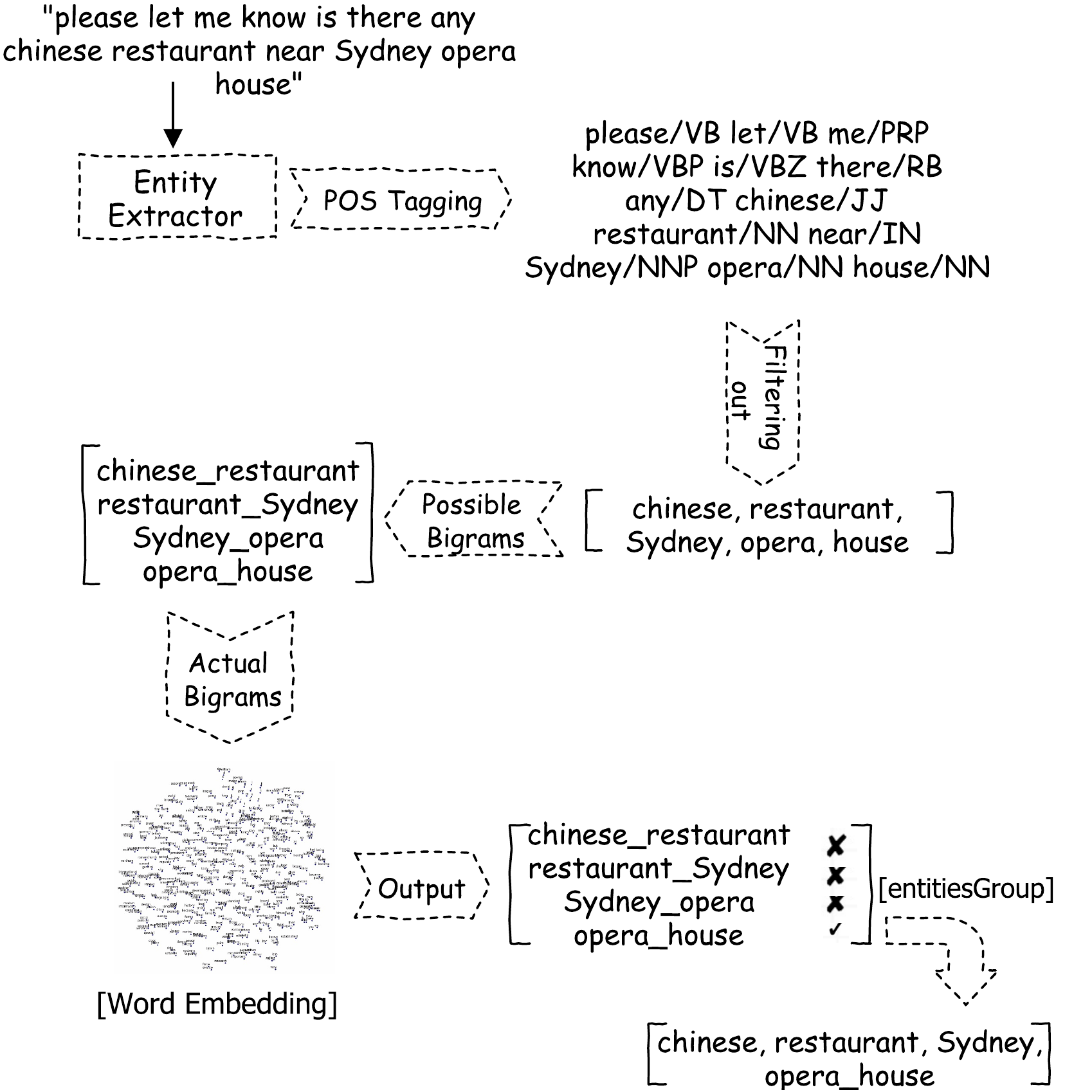}
\caption{Entity Extractor using Stanford POS Tagger and Word Embedding}
\vspace{-0.33cm}
\label{fig:entity-extractor}
\end{figure}

\begin{figure*}
\centering
\includegraphics[scale=0.57]{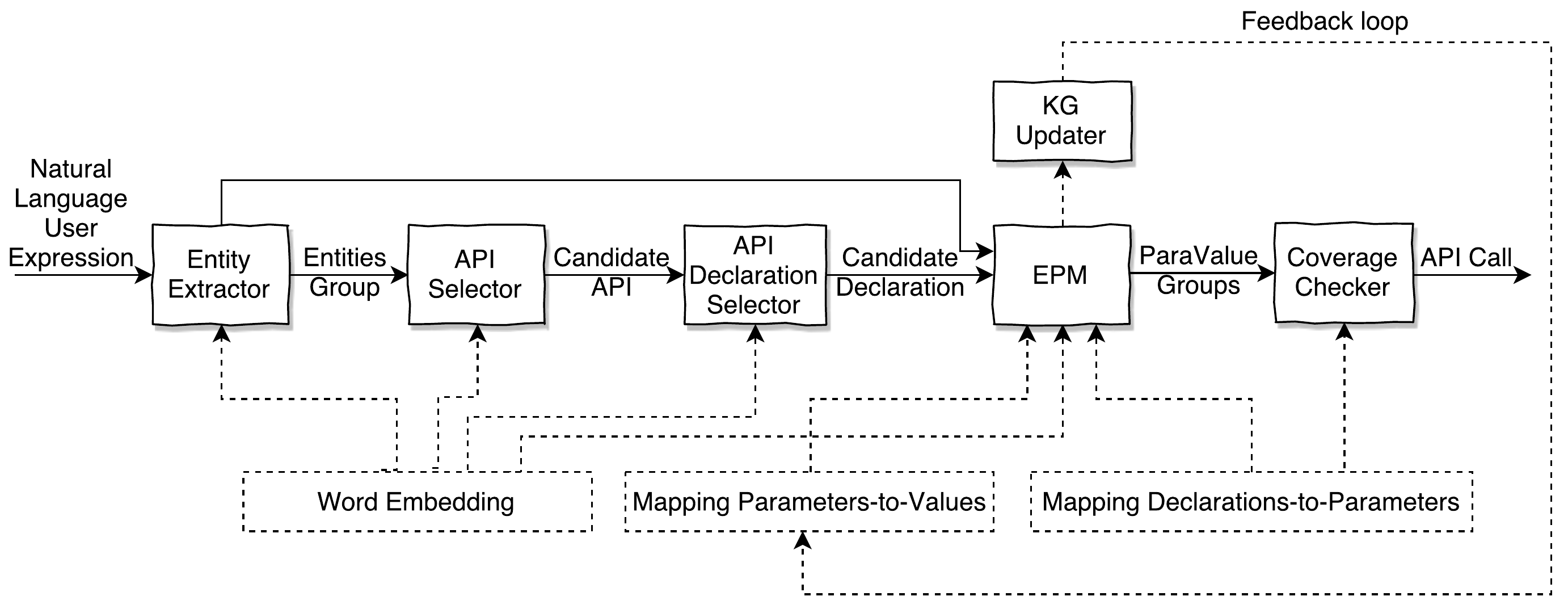}
\vspace{-0.2cm}
\caption{Synthesis Process -- Natural Language User Expressions into API Invocations}
\label{fig:synthesisapi}
\end{figure*}

Figure \ref{fig:synthesisapi} illustrates the functional architecture of these components in orchestration. We now describe in more detail, the role and individual functionality of each of these components:

\vspace{0.2cm}\noindent\textbf{Entity Extractor.} This component takes a user expression and decomposes into a set of entities; which are then classified into nouns, adjectives and proper nouns with the help of Stanford POS tagger \cite{stancorenlp-paper}. This filters all words that are irrelevant (e.g., stop words like \textit{the} and \textit{at}). Secondly, this component derives bigrams (e.g., \textit{ice cream} and \textit{New York}) with the help of our word embedding model (Figure \ref{fig:entity-extractor}).

\noindent
\textbf{API Selector.} Upon performing entity extraction from the inputted user expression, this provides the first step towards understanding the basic intent. Accordingly, we can use this information to select APIs that could potentially be a match in satisfying the user's request. We can match with APIs based on semantic similarity: this means, using the extracted entity information (i.e. nouns, verbs, as well as complex analysis derived from bigrams, see Figure \ref{fig:entity-extractor}), we can match with APIs in the KG by comparing with their associated tags, parameter names and values. It is important to note finding the best API match will require additional layers of analysis, namely examining the declarations of APIs and checking if their parameter signatures are covered. This will therefore be dealt with at subsequent steps (as we will describe in more detail below). However, the purpose of this stage is to perform an early filtration and produce a set of viable APIs.

\vspace{0.2cm}\noindent\textbf{API Declaration Selector.} Now that we have a viable set of APIs, the next step is identifying the best matched declarations. We may recall, declaration curated in the KG are linked with ``sample expressions''. Accordingly, we can perform semantic similarity analysis between the inputted user expression and the set of sample expressions stored for each declaration. To implement this, we again leverage our word embedding model to compute semantic similarity. We compute the vector representation of both the inputted user expression and each of the sample expressions as follows:

\begin{equation} \label{eq:1}
\begin{split}
& \overrightarrow{S}=\sum_{i=1}^{n}\overrightarrow{S_{w_i}} \quad where \: \overrightarrow{S_{w_i}} \in {W\!E} \\
& \overrightarrow{T^k}=\sum_{i=1}^{n}\overrightarrow{T^k_{w_i}} \quad where \: \overrightarrow{T^k_{w_i}} \in {W\!E}
\end{split}
\end{equation}

\noindent
where $\overrightarrow{S}$ and $\overrightarrow{T^k}$ are the vector representation of the user expression and a sample expression from the API Knowledge Graph respectively; $\overrightarrow{S_{w_i}}$ and $\overrightarrow{T^k_{w_i}}$ are the vector representation of a single word within each expression. $WE$ is the vector space defined by the Word Embedding model.

The actual semantic similarity of $\overrightarrow{S}$ and $\overrightarrow{T^k}$ is computed using the cosine similarity metric \cite{Jurafsky:2009:SLP:1214993}. Using this metric, the \emph{API Declaration Selector} chooses the declaration that has the highest score for $similarity$, which is equated as follows:

\begin{equation} \label{eq:methodstar}
\begin{split}
& expr^* = \operatorname*{arg\,max}_{\overrightarrow{T^k_i}} \big[similarity(\overrightarrow{S}, \overrightarrow{T^k_i})\big]\\
& dec* = declaration(expr^*)
\end{split}
\end{equation}

\noindent
where $expr^*$ is the sample expression from the API Knowledge Graph that has the highest similarity with the user expression $\overrightarrow{S}$, and $dec^*$ is the declaration that is associated to the sample expression. The declaration is obtained through the function $declaration$ above that takes as input an expression and returns an API declaration.
 
\vspace{0.2cm}\noindent\textbf{Entity-Parameter Mapper(EPM).} We are now ready to map extracted entities to the parameters of the chosen declaration. Let's return to our earlier example for illustration. Consider we have a user expression that states, \textit{``Is there any Chinese restaurant near Sydney Opera House''}, and the selected API declaration is \texttt{api.yelp.com/search?term=[term]\&location= [location]}. The job of the Entity-Parameter Mapper is to recognize that \textit{Chinese restaurant} should be mapped to the parameter \textit{term}, and that \textit{Sydney Opera House} to \textit{location}. 

In order to determine which entity should be mapped to which parameter, we once again apply the Word Embedding model. We compute for each parameter of the selected API declaration, the semantic similarity with stored parameter values for that declaration. (We may recall these stored values are acquired based on the methods discussed at Section~\ref{sec:knowledge-aq}). Therefore, more specifically in this example, if we compare the input entity \textit{Chinese}, with all stored parameter values of the selected API declaration, we could conclude the keyword \textit{Chinese} is probably a value of the \textit{term} parameter because of its semantic similarity with e.g. 'French', which is also a value of the \textit{term} parameter as stored in the Knowledge Graph.

The output of the EPM component is \textit{paraValue} matrix. It contains parameter-value pairs (i.e. mapped entities to parameters), together with confidence ratio for each. We also use a component named \textit{KG updater} to store new values into the knowledge graph as users are conversing with the bot. This component only store new values provided the confidence ratio above a certain threshold. At this stage, we set it to $0.40$. However, this may considered an initial threshold for experimentation purposes. We believe, as increasingly more knowledge is gained, it will become clearer to understand the right balance, and this figure will be more precise.

\vspace{0.2cm}\noindent\textbf{Coverage Checker.} This component is responsible for performing two main tasks: (i) Firstly, to choose the best declaration that has the maximum number of fulfilled required parameters amongst candidate API declarations; (ii) Secondly, to make sure that all the required parameters of the selected declaration are fulfilled before calling the API. Accordingly, the \emph{Coverage Checker} component takes two inputs: the candidate API declarations; and secondly, the \textit{paraValue} matrix that was produced by the previous component. 

We determine coverage for each viable declaration by computing the following:

\begin{equation} \label{eq:coverage}
\begin{split}
& coverage(dec_k) =\frac{\sum_{i=1}^N mapping(p_i)}{N} \\
& mapping(p_i)=\begin{cases}1 & p_i  \textrm{ is mapped to a value}
\\0 & p_i  \textrm{ is not mapped to a value}\end{cases}
\end{split}
\end{equation}

\vspace{0.2cm}
\noindent
where $coverage(dec_k)$ represents the ratio of parameters of the declaration $dec_k$ with a value assigned to it, $mapping(p_i) \in \{0,1\}$ indicates whether or not a parameter has been mapped to a value, and $N$ is the total count of required parameters for the API declaration. When this component is performed during execution, we require that the $coverage(dec_k)$ should be equal to $1$. This implies, all required parameters of the API declaration are fulfilled. When this happens, a call to the API can be performed and the results returned as a JSON object for processing of the final results and display to end-user. If $coverage(dec_k)$ is not equal to $1$, then \emph{Coverage Checker} notifies the end-user, and tries to obtain the necessary feedback about parameters for which no suitable value could be found.

In some cases, we may want to choose one declaration amongst a set (e.g., many declarations appear too similar, or if used during development mode, the bot developer may simply wants help to compare various declarations). Accordingly, the best declaration is picked based on equation (\ref{eq:decstar}):

\begin{equation} \label{eq:decstar}
dec^* = \operatorname*{arg\,max}_{dec_k} \big[coverage(dec_k)\big]
\end{equation}

\noindent
where $dec^*$ represents the declaration that scored the highest $coverage$ metric, as defined in equation (\ref{eq:coverage}). As a result of this, the Coverage Checker outputs the best API declaration.

\section{Related Work and Discussions}
There is a considerable body of research conducted on spoken dialog systems. Some works used crowds in areas like writing and editing \cite{Bernstein:2011:CTS:2047196.2047201}, image description and interpretation \cite{Bigham:2010:VNR:1866029.1866080}. Some others worked on taking advantage of the crowd to empower the dialog system abilities (answer to questions in several domains with various styles) by creating API calls \cite{huang2015guardian}. \emph{BotBase} takes inspiration from the latter work for building API calls dynamically, albeit by using machine learning and knowledge graph. 

Besides, there are some works focused on API recommendation techniques which are relevant to our work. RACK \cite{rahman2016rack} translates a natural language query into a list of relevant APIs by mapping keywords to APIs, Portfolio \cite{McMillan:2011:PFR:1985793.1985809} recommends relevant API methods for a given natural language query by employing indexing and graph based algorithms, proposed technique in \cite{chan2012searching} recommends a graph of API methods from precise textual phrase matching.

Furthermore, improvements in natural language processing techniques as well as bot development approaches have led to a renewal in bots. Most recent efforts in natural language processing have focused on translating user expressions into program codes \cite{Desaiprogram} \cite{Gulwaninlyze}. For bot development side, rule-based approaches such as AIML, RiveScript, ChatScript, Pandorabots and Superscript provide scripting-based languages for describing service-user conversations as a rule set (e.g., if-then statements). Flow-based approaches such as Motion.ai, ChatFuel, Manychat, FlowXO and Sequel provide platforms to describe end-user conversations as a flowchart (e.g. sequence of options and actions). Hybrid approaches like Wit.ai, API.ai, Microsoft LUIS, Meya, Pullstring, Gupshup, Microsoft BotFramework, IBM Conversation Service and Amazon Lex provide platforms to describe service-user conversations as intents and natural language expressions. However, these approaches are forcing developers to handle almost all the parts (e.g. training bot, writing pieces of codes for each user intents, interacting with underlying backend services such as APIs, code invocation commands and programming libraries) to generate user responses. \emph{BotBase} benefits from techniques in hybrid platforms to learn possible values for API parameters and to enrich its API Knowledge Graph.

The idea of building an API Knowledge graph for \emph{BotBase} comes from Augur \cite{fast2016augur} which is a knowledge graph of human actions, activities and their relation with objects; KnowBot \cite{hixon2015learning} a question-answer system that builds a knowledge graph of concepts (in science) and their relations through conversational dialogs between user and system; Commandspace \cite{Adar:2014:CMR:2642918.2647395} a knowledge graph of user expressions and Adobe Photoshop application commands; and QF-Graph \cite{Fourney:2011:QGB:2047196.2047224} a mapping between user's vocabulary and system features. We use the same idea to construct API knowledge graph contains API, API declarations, parameters and values. Furthermore, \emph{BotBase} leverages concepts in previous effort on synthesizing Java language codes for a given input text by using PCFG model and mapping words and API declarations \cite{Gvero:2015:SJE:2814270.2814295}. \emph{BotBase} constructs mappings between API declarations to sample expressions and parameters to values in its knowledge graph.

\section{Limitations \& Future Work}
In this paper, we have unleashed the first generation of our vision towards an era of intelligent bots that learns from complex data sets and mimics the way of the human brain. At this stage, we make certain assumptions, we elaborate upon these limitations below together with possible solutions for each.

\vspace{0.2cm}\noindent
\textbf{Conversational Bots.}
As \textit{BotBase} is only within its early stages, it does not yet support conversational bots. At this stage we assume a stateless environment, where each question from the user has an answer from the bot, and the bot does not remember anything from the previous chat messages. Future works therefore requires a stateful conversation mechanism in \emph{BotBase}. 

\vspace{0.2cm}\noindent
\textbf{Process Workflows.}
One of the most significant limitations of our current work is the assumption that an action can be performed by one API declaration alone. While it is true, there are currently a plethora of APIs and large amounts of user requests could in fact be fulfilled by a single declaration. Nevertheless, in future work we intend to support executing dynamic process workflows (i.e. calling upon a series or combination of APIs) to fulfill a user request.

\vspace{0.2cm}\noindent
\textbf{Advanced Knowledge Mining.} The success of this proposed approach depends greatly on the quantity and accuracy of the knowledge graph. At present, we have acquired a sizable quantity of data using bootstrapping, crowdsourcing and general purpose Wikipedia corpus to train our word embedding model. This truly does supply an enormous amount of data, however in future we would benefit from gathering from other sources of data, no only formal datasets or corpuses.

\section{Conclusion}
In this paper, we provide a solution for dynamically synthesizing natural language user expressions into API invocations. To achieve this we have proposed a lightweight API Knowledge Graph that contain the required information about APIs. We enrich the KG with new APIs, as well as appending supplementary knowledge about existing APIs. Enrichment is achieved with the intelligence of deep-learning (Word Embedding model trained by Wikipedia corpus) in conjunction with crowdsourcing approaches. In the synthesis process, \emph{BotBase} benefits from the API KG and the word embedding model to determine which API to invoke based on user expressions. 


\section*{Acknowledgment}
We Acknowledge Data to Decisions CRC for funding scholarship on Query Answering and Predictive Techniques for Analyst Tasks in End-User Big Data Analytic.



%



\bibliographystyle{IEEEtran}
\bibliography{IEEEabrv,bare_conf}
\end{document}